\documentstyle[prl,aps,floats,epsf,psfig,colordvi]{revtex}
\draft
\newcommand{\nub}{\overline{\nu}}
\newcommand{\cbar}{\overline{c}}

\def\nim#1#2#3  {{\it Nucl. Instr. Meth.} {\bf#1}, #2 (#3). }
\def\np#1#2#3   {{\it Nucl. Phys.} {\bf#1}, #2 (#3). }
\def\pcps#1#2#3 {{\it Proc. Cam. Phil. Soc.} {\bf#1}, #2 (#3). }
\def\pl#1#2#3   {{\it Phys. Lett.} {\bf#1}, #2 (#3). }
\def\plc#1#2#3   {{\it Phys. Lett.} {\bf#1}, #2 (#3); }
\def\prep#1#2#3 {{\it Phys. Rep.} {\bf#1}, #2 (#3). }
\def\prev#1#2#3 {{\it Phys. Rev.} {\bf#1}, #2 (#3). }
\def\prl#1#2#3  {{\it Phys. Rev. Lett.} {\bf#1}, #2 (#3). }
\def\prs#1#2#3  {{\it Proc. Roy. Soc.} {\bf#1}, #2 (#3). }
\def\rmp#1#2#3  {{\it Rev. Mod. Phys.} {\bf#1}, #2 (#3). }
\def\rpp#1#2#3  {{\it Rep. Prog. Phys.} {\bf#1}, #2 (#3). }
\def\zp#1#2#3   {{\it Z. Phys.} {\bf#1}, #2 (#3). }
\def\epj#1#2#3   {{\it Eur. Phys. Jour.} {\bf#1}, #2 (#3). }

\begin{document}

\wideabs{
\title{ Measurements of 
$F_2$ and $xF_3^{\nu}-xF_3^{\nub}$
from CCFR $\nu_\mu$-Fe and $\nub_\mu$-Fe 
 data in a physics model independent way}

\author{U.~K.~Yang,$^{7}$ T.~Adams,$^{4}$ A.~Alton,$^{4}$
C.~G.~Arroyo,$^{2}$ S.~Avvakumov,$^{7}$ L.~de~Barbaro,$^{5}$
P.~de~Barbaro,$^{7}$ A.~O.~Bazarko,$^{2}$ R.~H.~Bernstein,$^{3}$
A.~Bodek,$^{7}$ T.~Bolton,$^{4}$ J.~Brau,$^{6}$ D.~Buchholz,$^{5}$
H.~Budd,$^{7}$ L.~Bugel,$^{3}$ J.~Conrad,$^{2}$ R.~B.~Drucker,$^{6}$
B.~T.~Fleming,$^{2}$
J.~A.~Formaggio,$^{2}$ R.~Frey,$^{6}$ J.~Goldman,$^{4}$
M.~Goncharov,$^{4}$ D.~A.~Harris,$ ^{7} $ R.~A.~Johnson,$^{1}$
J.~H.~Kim,$^{2}$ B.~J.~King,$^{2}$ T.~Kinnel,$^{8}$
S.~Koutsoliotas,$^{2}$ M.~J.~Lamm,$^{3}$ W.~Marsh,$^{3}$
D.~Mason,$^{6}$ K.~S.~McFarland, $^{7}$ C.~McNulty,$^{2}$
S.~R.~Mishra,$^{2}$ D.~Naples,$^{4}$  P.~Nienaber,$^{3}$
A.~Romosan,$^{2}$ W.~K.~Sakumoto,$^{7}$ H.~Schellman,$^{5}$
F.~J.~Sciulli,$^{2}$ W.~G.~Seligman,$^{2}$ M.~H.~Shaevitz,$^{2}$
W.~H.~Smith,$^{8}$ P.~Spentzouris, $^{2}$ E.~G.~Stern,$^{2}$
N.~Suwonjandee,$^{1}$ A.~Vaitaitis,$^{2}$ M.~Vakili,$^{1}$   
J.~Yu,$^{3}$ G.~P.~Zeller,$^{5}$ and E.~D.~Zimmerman$^{2}$}

\address{( The CCFR/NuTeV Collaboration ) \\
$^{1}$ University of Cincinnati, Cincinnati, OH 45221 \\
$^{2}$ Columbia University, New York, NY 10027 \\
$^{3}$ Fermi National Accelerator Laboratory, Batavia, IL 60510 \\
$^{4}$ Kansas State University, Manhattan, KS 66506 \\
$^{5}$ Northwestern University, Evanston, IL 60208 \\
$^{6}$ University of Oregon, Eugene, OR 97403 \\
$^{7}$ University of Rochester, Rochester, NY 14627 \\
$^{8}$ University of Wisconsin, Madison, WI 53706\\ }

\maketitle
\begin{abstract}

We report on the extraction of  the structure functions
$F_2$ and  $\Delta xF_3 = xF_3^{\nu}-xF_3^{\nub}$
from CCFR  $\nu_\mu$-Fe and $\nub_\mu$-Fe differential cross sections.
The extraction is performed in a physics model independent (PMI) way.
This first measurement of  $\Delta xF_3$, which is useful in testing
models of heavy charm production, is higher
than current theoretical predictions.
The ratio of the $F_2$ (PMI) values 
measured in $\nu_\mu$ and $\mu$ scattering
is in agreement  (within 5\%) with
the predictions of Next-to-Leading-Order parton distribution functions
(NLO PDFS) using massive charm production schemes, thus
resolving the long-standing discrepancy between the two
sets of data.
\end{abstract}
\pacs{PACS numbers:12.38.Qk, 13.15.+g, 24.85.+p, 25.30.Pt 
{\it UR-1586, submitted to Phys. Rev. Lett.} }
\twocolumn
}



 Deep inelastic lepton-nucleon scattering experiments have been used
to determine the quark distributions in the nucleon.
However, the quark distributions determined from muon~\cite{NMC} 
and neutrino~\cite{SEL}
experiments were found to be different at small values of Bjorken $x$,
because of a disagreement in the extracted structure functions.
 In this Letter,
we report on a measurement of differential cross sections
and structure functions from CCFR $\nu_\mu$-Fe and $\nub_\mu$-Fe data.
The neutrino-muon difference is resolved by
extracting the $\nu_\mu$ structure functions in a
physics model independent (PMI)
way. We also report on the first measurement of
 $\Delta xF_3$ = $xF_3^{\nu}-xF_3^{\nub}$, which is
used to test models of heavy charm production.

The sum of $\nu_\mu$ and $\nub_\mu$
 differential cross sections 
for charged current interactions on an isoscalar target is related to the
structure functions as follows:
\begin{tabbing}
$F(\epsilon)$ \= $\equiv \left[\frac{d^2\sigma^{\nu }}{dxdy}+
\frac{d^2\sigma^{\overline \nu}}{dxdy} \right] 
 \frac {(1-\epsilon)\pi}{y^2G_F^2ME_\nu}$ \\
  \> $ = 2xF_1 [ 1+\epsilon R ] + \frac {y(1-y/2)}{1+(1-y)^2} \Delta xF_3 $. \hspace{0.7in} (1)
\end{tabbing}
Here $G_{F}$ is the Fermi weak coupling constant, $M$ is the nucleon
mass, $E_{\nu}$ is the incident neutrino energy, the scaling
variable $y=E_h/E_\nu$ is the fractional energy transferred to
the hadronic vertex, $E_h$ is the final state hadronic
energy, and $\epsilon\simeq2(1-y)/(1+(1-y)^2)$ is the polarization of 
the virtual $W$ boson.
The structure
function $2xF_1$ is expressed in terms of $F_2$
by $2xF_1(x,Q^2)=F_2(x,Q^2)\times
\frac{1+4M^2x^2/Q^2}{1+R(x,Q^2)}$, where $Q^2$ is the
square of the four-momentum transfer to the nucleon,
  $x=Q^2/2ME_h$ (the Bjorken scaling variable) is
the fractional momentum carried by the struck quark,
and $R=\frac{\sigma
_{L}}{\sigma _{T}} $ is the ratio of the cross-sections of
longitudinally- to transversely-polarized $W$ bosons.

A similar equation for the case of muon scattering
 relates the
cross sections to the structure functions. However, there
are significant differences originating from the scattering
on strange ($s$) and charm ($c$) quarks.
The $\Delta xF_3$ term, which in leading order
 $\simeq 4x(s-c)$, is not present in the $\mu$ scattering case.
In addition, in a charged current $\nu_\mu$ interaction involving
$s$ (or $\cbar$) quarks, there is a threshold suppression originating
from the production of heavy $c$ quarks in the final state.
For $\mu$ scattering, while there is no suppression for scattering
from $s$ quarks, there is larger suppression when scattering
from $c$ quarks since there are two heavy
 quarks ($c$ and $\cbar$) in the final state.  

In previous analyses~\cite{SEL} of
 $\nu_\mu$ data,
light flavor universal
physics model dependent (PMD) structure functions were extracted
by applying a slow rescaling correction to correct for the charm
mass suppression in the final state. In addition, 
the $\Delta xF_3$ term (used as input in the extraction)
was calculated from a leading order charm production model. Recent
calculations~\cite{sch_dep,MFS,vfs} indicate that there are large
theoretical uncertainties in the charm production modeling for both
$\Delta xF_3$ and the slow rescaling corrections. Therefore, in the
new analysis reported here, slow rescaling corrections are not applied,
 and $\Delta xF_3$ and $F_2$
are extracted from two-parameter fits to the data.
We compare the values of $\Delta xF_3$
to various charm production models.
The extracted physics model independent (PMI) values for  $F_2^{\nu}$
are then compared  with $F_2^{\mu}$
within the framework of NLO models for charm production.


The CCFR experiment  collected  data
using the Fermilab Tevatron Quad-Triplet wide-band  $\nu_\mu$ and $\nub_\mu$
beam. The CCFR detector~\cite{CALIB} consists of a
 steel-scintillator target calorimeter
instrumented with drift chambers, followed by a toroidally 
magnetized muon spectrometer.
The hadron energy resolution is
$\Delta E_h/E_h = 0.85/\sqrt{E_h}$(GeV), and the muon momentum resolution is 
$\Delta p_\mu/p_\mu = 0.11$. By measuring the hadronic energy ($E_h$), muon
momentum ($p_\mu$), and muon angle ($\theta_\mu$), we construct
three independent kinematic variables $x$, $Q^2$, and $y$.
The relative flux at different energies, obtained from the events 
with low hadron energy ($E_h < 20$ GeV), is normalized so that
the neutrino total cross section equals the world average $\sigma^{\nu N}/E=
(0.677\pm0.014)\times10^{-38}$ cm$^2$/GeV and $\sigma^{\overline{\nu} N}
/\sigma^{\nu N}=0.499\pm0.005$~\cite{SEL}. 
 After fiducial and kinematic cuts ($p_{\mu}>15$ GeV, 
$\theta_{\mu} <0.150$, $E_h > 10$ GeV, and 30 GeV $<E_{\nu}<$ 360 GeV),
the data sample 
consists of
1,030,000 $\nu_{\mu}$ and 179,000 $\nub_{\mu}$
events. Dimuon events are removed because of the ambiguous identification 
of the leading muon for high-$y$ events. 

\begin{figure}[t]
\centerline{\psfig{figure=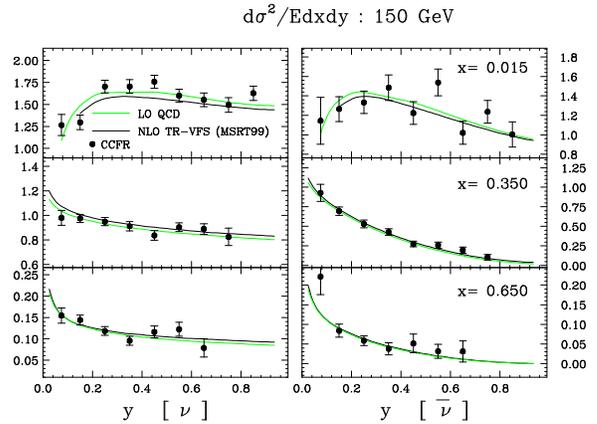,width=3.0in}}
\caption{Typical raw differential cross sections at $E_\nu=150$ GeV
(both statistical and systematic errors are included).
 The CCFR data are in
good agreement with the NLO TR-VFS QCD calculation using MRST99 PDFs
(dashed line).
The solid line is a CCFR LO QCD inspired fit.}
\label{fig:diff}
\end{figure}

The raw differential cross sections per nucleon on iron
 are determined in bins of $x$, $y$, 
and $E_{\nu}$ ($0.01 < x < 0.65$, $0.05<y<0.95$, and $30< E_\nu <360$ GeV). 
Figure \ref{fig:diff} shows typical differential cross sections 
at $E_\nu=150$ GeV (complete tables are available~\cite{WWW}).
For all energies, the cross sections are in 
good agreement with NLO PDFs (with massive charm
production schemes e.g., MRST99~\cite{MRST} or CTEQ4HQ~\cite{CTEQ4HQ}).
The dashed lines shows the predictions
from the Thorne and Roberts Variable Flavor Scheme (TR-VFS)~\cite{MFS} 
QCD calculation
 using MRST99 extended~\cite{ext} PDFs.
This calculation includes an improved treatment of massive charm
production. The QCD predictions, which are on free neutrons and
protons, are corrected for nuclear~\cite{SEL},
 higher twist~\cite{dupaper,YANGR},
and radiative effects~\cite{BARDIN}.
Also shown are
the predictions from a CCFR leading order (LO) QCD inspired fit
used for calculation of acceptance and resolution smearing corrections.
As expected from QCD,
 the CCFR cross section data exhibit a quadratic $y$ dependence at small $x$ 
for $\nu_\mu$ and $\nub_\mu$, and a flat $y$ distribution at high $x$ 
for $\nu_\mu$.
%

Next, the raw cross sections are corrected for electroweak radiative
effects~\cite{BARDIN}, the $W$ boson propagator, and for the
 5.67\% non-isoscalar excess 
of neutrons over protons in iron (only important at high $x$).
Values of $\Delta xF_3$ and $F_2$ are extracted from the sums of
the corrected $\nu_\mu$-Fe and $\nub_\mu$-Fe 
differential cross sections according to Eq.~(1). 
However, it is challenging to fit $\Delta xF_3$, $R$, and $2xF_1$ 
using the $y$ distribution at a given $x$ and $Q^2$
because of the  strong correlation 
between the $\Delta xF_3$ and $R$ terms, unless the full range of
$y$ is covered by the data.
Covering this range 
(especially the high $y$ region) is difficult
because of the low acceptance. Therefore, we restrict
the analysis to two-parameter fits.

Our strategy is  to fit $\Delta xF_3$ and $F_2$ (or equivalently $2xF_1$)
for $x<0.1$ where the $\Delta xF_3$ contribution is relatively large, while
constraining $R$ using the
$R_{world}^{\mu/e}$~\cite{RWORLD} QCD-inspired empirical fit
to all available electron and muon scattering data.
The $R_{world}^{\mu/e}$ fit is also in good agreement
with NMC $R^\mu$ data~\cite{NMC} at low $x$, and
with the most recent theoretical prediction~\cite{YANGR}
for $R$ (a NNLO QCD calculation including target mass effects).
For $x<0.1$,  $R$ in neutrino scattering is expected to be somewhat larger
than $R$ for muon scattering because 
of the production of massive charm quarks
in the final state.  A correction for this
difference is applied to $R_{world}^{\mu/e}$ using a
 leading order slow rescaling model
to obtain an effective $R$ for neutrino scattering, $R_{eff}^{\nu}$. The
difference between  $R_{world}^{\mu/e}$ and $R_{eff}^{\nu}$
 is used as a systematic
error.  Because of the positive correlation between
$R$ and $\Delta xF_3$, the extracted values of $F_2$ are rather insensitive
to the input $R$. 
In contrast, the extracted values of $\Delta xF_3$
are sensitive to the assumed value of $R$, which is reflected in
a larger systematic error. 
The values of $\Delta xF_3$ are sensitive to the energy 
dependence of the neutrino flux ($\sim$ $y$ dependence),
 but are insensitive to the absolute normalization.
The uncertainty on the flux shape is estimated by using the 
constraint that $F_2$
and $xF_3$ should be flat over $y$ (or $E_{\nu}$) for each $x$ and $Q^2$ bin.

\begin{figure}[t]
\centerline{\psfig{figure=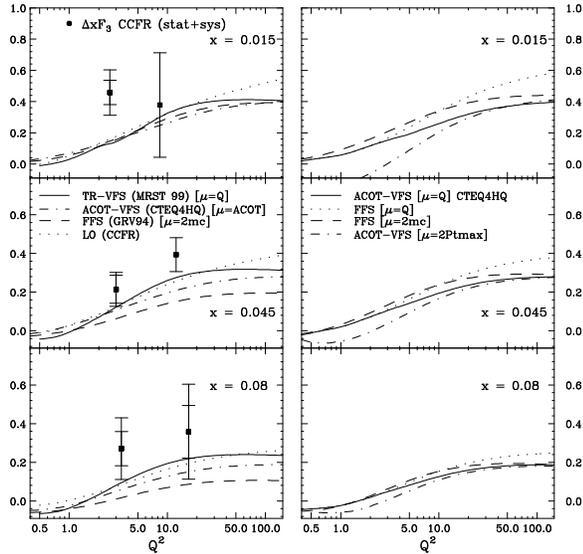,width=3.0in}}
\caption{ $\Delta xF_3$ data as a function of $x$ 
compared with various schemes for massive charm production:(left) 
TR-VFS(MRST99),
ACOT-VFS(CTEQ4HQ), FFS(GRV94), and the CCFR-LO (a leading order model with a slow rescaling correction);
(right) Sensitivity of the theoretical calculations
to the choice of scale.}
\label{fig:dxf3}
\end{figure}

\begin{figure}[t]
\centerline{\psfig{figure=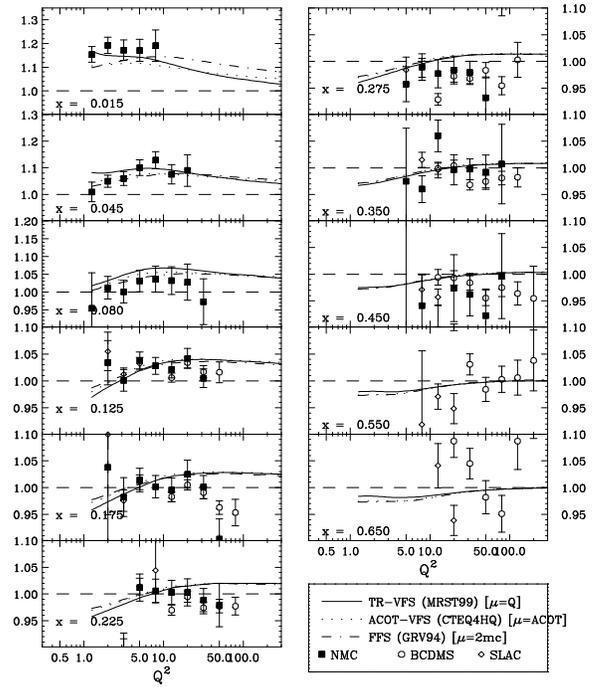,width=3.0in}}
\caption{
The ratio of
$F_2^{\nu}$ (PMI) data divided by $F_2^{\mu}$ (NMC or BCDMS)
or $F_2^e$ (SLAC). 
Both statistical and systematic errors are included.
Also shown are the predictions of the
TR-VFS (MRST99), ACOT-VFS (CTEQ4HQ) and FFS (GRV94) heavy flavor calculations.}
\label{fig:f2}
\end{figure}
Because of the limited statistics, we use 
large bins in $Q^2$ in the extraction of $\Delta xF_3$ with bin centering
corrections applied using the NLO
TR-VFS calculation~\cite{MFS} with the MRST99 PDFs.
Figure~\ref{fig:dxf3} (left) shows the extracted
values of $\Delta xF_3$ as a function of $x$, including both statistical
and systematic errors, compared to
various theoretical methods for modeling
heavy charm production within a QCD framework. 
The three-flavor Fixed Flavor Scheme (FFS)~\cite{FFS} assumes 
that there is no intrinsic charm 
in the nucleon, and that all scattering from
$c$ quarks occurs via the gluon-fusion diagram.  The concept behind the
Variable Flavor Scheme (VFS) proposed by ACOT~\cite{vfs,ACOT} is that at low
scale, $\mu$, one uses the three-flavor FFS scheme, while above some scale,
an intrinsic charm sea (which is evolved from zero) is introduced. 
The concept of
the TR-VFS scheme~\cite{MFS} is similar, except that at
intermediate scale it
interpolates smoothly between the two regions. Both the FFS and VFS schemes
have been implemented by  KLS~\cite{sch_dep}, ACOT, and
Kretzer~\cite{kretzer}. The last two implementations agree with each
other, but not with KLS (there was a mistake in the
KLS calculation)~\cite{private}.

Shown are the predictions from the
TR-VFS scheme (implemented with MRST99 PDFs and the suggested scale
 $\mu=Q$) and along with the predictions from two other
other NLO calculations,
ACOT-VFS (implemented with CTEQ4HQ PDFs
and their recent  suggested scale $\mu = m_c$ for $Q<m_c$, and
$\mu^2=m_c{^2}+0.5Q^2(1-m_c{^2}/Q^2)^n$ for $Q<m_c$ with $n=2$
~\cite{vfs}), and 
the FFS (implemented with the GRV94~\cite{GRV94}  PDFs and
 their recommended scale $\mu = 2m_c$).
Also shown are the predictions 
from the leading order QCD fit to the CCFR dimuon~\cite{dimu}
data.

Figure~\ref{fig:dxf3} (right) shows the sensitivity to
the choice of scale. For example,
the data do not favor the choice of scale, $\mu=2Pt_{max}$
in the ACOT-VFS calculation with CTEQ4HQ PDFs.
This high scale (originally
suggested by ACOT and used in the CCFR dimuon
analysis~\cite{dimu}) implies that the calculation 
is in the four-flavor region even at $x=0.015$ and $Q^2=1.0$ GeV$^2$
(and yields large negative result).
With reasonable choices of scale, all the theoretical models
yield similar results. However,
at low $Q^2$,  our $\Delta xF_3$ data is higher than all 
of the theoretical models.
The difference between data and theory 
may be due to an underestimate
of the strange sea at low $Q^2$, or from missing NNLO terms.
The question of the strange sea would be addresed 
by a  global NLO analysis which combines the neutrino data 
for dimuons, $\Delta xF_3$ and $F_2^\nu$, with $F_2^{\mu, e}$.

As discussed above, values of $F_2$ for $x<0.1$ are extracted from
two-parameter fits to the $y$ distributions.
In the $x>0.1$ region,
the contribution from  $\Delta xF_3$ is small and the extracted
values of $F_2$ are insensitive to $\Delta xF_3$. Therefore,
we extract values of $F_2$ with an input value of $R$
and with $\Delta xF_3$ constrained to the TR-VFS (MRST99) predictions. 
%
%
Fig.~\ref{fig:f2} shows
the ratio of our $F_2^{\nu}$ (PMI) measurements divided by
$(18/5)F_2^{\mu}$ (NMC~\cite{NMC} or BCDMS~\cite{BCDMSF2}) or
$(18/5)F_2^{e}$ (SLAC~\cite{SLACF2})  measurements~\cite{bin-corr}.
The overall normalization errors of 2\% (CCFR), and 2.5\% (NMC) are not shown.
Within 5\%, the ratio  is in agreement with the predictions
of the TR-VFS (MRST99), ACOT-VFS (CTEQ4HQ),
and FFS (GRV94) calculations~\cite{note}.

In the calculation of the theoretical
 predictions, we have also included corrections for nuclear
effects~\cite{SEL}. 
As mentioned earlier, 
the extracted values of $F_2$ from the two-parameter fits 
are insensitive to $R$.
For example, 
if we 
perform simultaneous two-parameter fits to $F_2$ and $R$ (while
keeping $\Delta xF_3$
fixed to the TR-VFS (MRST99) values), the extracted $R$ values
at $x=0.01$ are
smaller than $R_{eff}^\nu$, 
but $F_2$ changes by only $2\sim 3\%$.

\begin{figure}[t]
\centerline{\psfig{figure=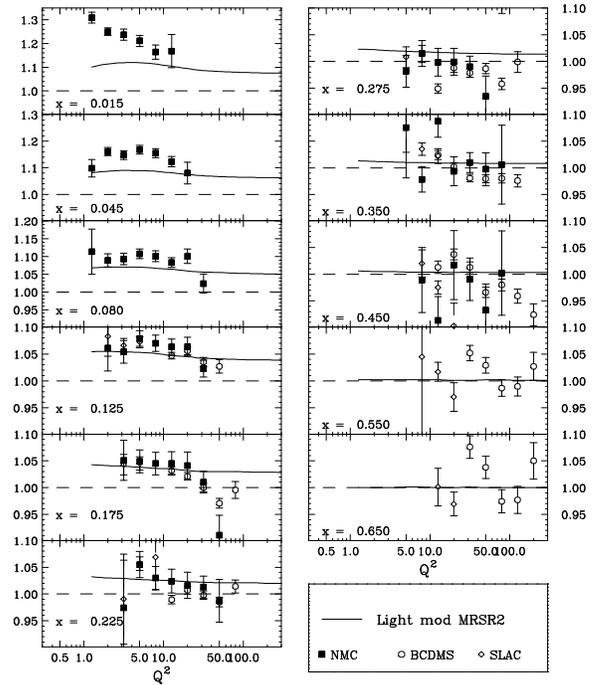,width=3.0in}}
\caption{
The ratio of the previous
$F_2^{\nu}$ (PMD) data divided by $(18/5)F_2^{\mu}$ (NMC or BCDMS)
or $(18/5)F_2^e$ (SLAC). Shown are the predictions of the
 MRSR2 light-flavor PDFs (the curves with CTEQ4M
are very similar).
}
\label{fig:f2old}
\end{figure}

In the previous analysis~\cite{SEL}
 of the CCFR data, the ratio of extracted values of 
$F_2^{\nu}$ (PMD) data divided by $(18/5)F_2^{\mu}$ (NMC) 
 at the lowest  $x=0.015$ and $Q^2$ bin
were $20\%$ higher than the predictions of the light-flavor PDFs such as
MRSR2~\cite{MRSR2} or CTEQ4M~\cite{CTEQ4HQ} (see Fig.~\ref{fig:f2old}  ). 
 About $10\%$ of the difference originates from having used a leading
order model for $\Delta xF_3$ versus using our new measurement.
 Another $6\%$ originates
from having used the leading order slow rescaling corrections, instead of
using 
 NLO massive charm production models. The remaining
3\% originates from improved modeling of the low $Q^2$ PDFs (which
changes the radiative corrections and the overall absolute normalization
to the total neutrino cross sections).
For higher $Q^2$ at $x=0.015$,
and for the next two  $x$ bins ($x=0.045$ and $0.08$), the smaller
difference between the PMI and PMD results is due to equal contributions
from the $\Delta xF_3$ and the difference in the slow rescaling corrections.
 For the higher $x$  bins ($x>0.1$), the contribution
of $\Delta xF_3$ is small, and the slow rescaling corrections
 in the leading
order model are the same as those with the
NLO theories. Therefore, the NMC
and CCFR data are in agreement at large $x$
whether  PMI or PMD structure functions are used in the comparison.

%


In conclusion, the ratio of $F_2$ (PMI) values
measured in neutrino-iron and muon-deuterium scattering
are in agreement with the 
predictions of Next-to-Leading-Order PDFs (using
massive charm production schemes),
 thus
resolving the long-standing discrepancy between the two
sets of data.
The first measurement of $\Delta xF_3$ 
is higher than current theoretical predictions.


\begin{references}
\bibitem{bin-corr}
An empirical fit~\cite{NMC} to the SLAC, BCDMS, NMC $F_2^d$ data 
is used for bin-center correction to these data.
\end{references}
\end{document}